\documentclass[%
reprint,
amsmath,amssymb,
prd,
]{revtex4-2}

\usepackage{graphicx}
\usepackage{dcolumn}
\usepackage{bm}
\usepackage{color}
\usepackage{lipsum}

\begin{document}

\preprint{APS/123-QED}

\title{Discovering a new well: Decaying dark matter with profile likelihoods}

\author{Emil Brinch Holm$^1$}
\email{Corresponding author: ebholm@phys.au.dk}
\author{Laura Herold$^2$}
\author{Steen Hannestad$^1$}
\author{Andreas Nygaard$^1$}
\author{Thomas Tram$^1$}

\affiliation{
	\vspace{0.2cm}
	$^1$Department of Physics and Astronomy, Aarhus University, DK-8000 Aarhus C, Denmark
}
\affiliation{
	$^2$Max-Planck-Institut für Astrophysik, Karl-Schwarzschild-Str. 1, 85748 Garching, Germany
}

\date{\today}

\begin{abstract}
A large number of studies, all using Bayesian parameter inference from Markov Chain Monte Carlo methods, have constrained the presence of a decaying dark matter component. All such studies find a strong preference for either very long-lived or very short-lived dark matter. However, in this letter, we demonstrate that this preference is due to parameter volume effects that drive the model towards the standard $\Lambda$CDM model, which is known to provide a good fit to most observational data. 

Using profile likelihoods, which are free from volume effects, we instead find that the best-fitting parameters are associated with an intermediate regime where around $3 \%$ of cold dark matter decays just prior to recombination. With two additional parameters, the model yields an overall preference over the $\Lambda$CDM model of $\Delta \chi^2 \approx -2.8$ with \textit{Planck} and BAO and $\Delta \chi^2 \approx -7.8$ with the SH0ES $H_0$ measurement, while only slightly alleviating the $H_0$ tension. Ultimately, our results reveal that decaying dark matter is more viable than previously assumed, and illustrate the dangers of relying exclusively on Bayesian parameter inference when analysing extensions to the $\Lambda$CDM model. 
\end{abstract}

\maketitle

\section{\label{sec:level1}Introduction}

The current concordance model of cosmology, $\Lambda$CDM, faces several tensions with observational data, arguably the most pressing of which is the $\sim 4 \sigma$ discrepancy between the value of the Hubble constant $H_0$ as measured from local supernovae~\cite{Riess:2021jrx} and the one inferred from the cosmic microwave background~\cite{Planck_col_2020}. In response, numerous extensions of $\Lambda$CDM are currently being proposed. Common to all of these is that they must reduce to $\Lambda$CDM in some region of their parameter space. However, in such a limit, all other parameters of the extension will become unconstrained, leading to an exaggerated emphasis on the $\Lambda$CDM values of parameter space once they are marginalized over during Bayesian parameter estimation. Such \emph{volume effects} can therefore greatly mislead conclusions about parameter constraints and the viability of the models, and developing techniques to detect these is therefore crucial to understanding many of the currently popular extended cosmological models.

One popular extension allows a fraction of the cold dark matter to decay to invisible radiation. Studies using Bayesian methods show that CMB data either prefers very short-lived or very long-lived decays, excluding any intermediary lifetimes. Recent studies have shown the short-lived class of models leads to a slight reduction in the Hubble tension~\cite{Holm:2022eqq, Nygaard:2020sow}, and other studies have shown that if the decay product is allowed to be massive, the model can greatly alleviate the tension between CMB and weak lensing measurements of the amplitude of matter fluctuations on an $8h^{-1}$ Mpc$^{-1}$ scale~\cite{Simon:2022ftd, FrancoAbellan:2021sxk}, known as the $\sigma_8$ tension. Decaying dark matter is therefore one of only a few models that are both physically well-motivated and may simultaneously address both the Hubble and $\sigma_8$ tensions. However, since the model converges to $\Lambda$CDM for long lifetimes and a model of dark radiation for small lifetimes, the preference for either of these limits could be driven by volume effects. 

In this letter, we present the first (to our knowledge) full frequentist analysis of cold dark matter decaying to dark radiation and show that a significantly different conclusion is reached compared to that of the Bayesian analyses. We constrain the cosmological parameters with profile likelihoods~\cite{Feldman_1998, Planck2014_PL}, which have recently seen renewed interest as tools to study volume effects~\cite{Herold:2021ksg, Gomez-Valent:2022hkb, Campeti:2022, Reeves:2022aoi, Herold:2022iib}, and instead of the one-sided MCMC bounds, we find a preference for around $3\textendash4\%$ of cold dark matter decaying between matter-radiation equality and recombination. Ultimately, our work clearly demonstrates the difference in conclusions reached by Bayesian analysis using Markov Chain Monte Carlo (MCMC) and frequentist analysis using profile likelihoods, thereby stressing the importance of using both methods complementarily for the future assessment of extensions to the $\Lambda$CDM cosmology.

\section{Decaying dark matter}\label{sec:level2}
Decaying dark matter generically refers to a family of models, the phenomenologically simplest\textemdash and most studied\textemdash model of which has a cold dark matter particle decaying into massless non-interacting particles dubbed dark radiation (DR). We will refer to this as the DCDM model, and assume that a fraction
$f_\text{dcdm}=\omega_\text{dcdm}^\text{ini}/(\omega_\text{dcdm}^\text{ini} + \omega_\text{cdm})$ of the cold dark matter decays with a decay constant $\Gamma$, where $\omega_\text{dcdm}^\text{ini}$ denotes the density parameter that the DCDM would have had today if it did not decay~\cite{Audren2014}. The possible production of additional radiation prior to recombination can lead to increased values of $H_0$, and MCMC analyses assert that the model alleviates the $H_0$ by $\sim 1\sigma$~\cite{Nygaard:2020sow, Holm:2022eqq, Simon:2022ftd}. In addition, we note that this particular model does not alleviate the $\sigma_8$ tension since it does not include massive decay products. For more details on the modelling aspects, we refer the reader to one of the many previous works on decaying cold dark matter~\cite{Audren2014, Poulin:2016nat, Nygaard:2020sow, Alvi:2022aam, DES:2020mpv, Pandey:2019plg}.

The DCDM model is well-constrained by high-$\ell$ CMB data due to reduced small-scale anisotropies from the massless decay products~\cite{Hou_2012}. In particular, references~\cite{Poulin:2016nat, Nygaard:2020sow} recognize two independently favoured regimes of long- or short-lived cold dark matter components, respectively, but find that the intermediate regimes are disfavoured. Only a few direct constraints on the decay constant $\Gamma$ have been derived in the general case where $f_\mathrm{dcdm} \neq 1$ due to the non-trivial correlation with $f_\mathrm{dcdm}$. Reference~\cite{Nygaard:2020sow} derive the strongest combined upper bound $\Gamma f_\mathrm{dcdm} < 3.78\times 10^4 \text{ Gyr}^{-1}$ in the short-lived regime, which is greatly relaxed in the long-lived regime, while references~\cite{Poulin:2016nat, Nygaard:2020sow, Schoneberg:2021qvd, Alvi:2022aam, Holm:2022eqq} all find posterior distributions in $\Gamma$ that are either bounded from above or below. In a similar manner, all previous MCMC analyses give upper bounds on the DCDM abundance with MAP estimates usually in the $\Lambda$CDM limit. Reference~\cite{DES:2020mpv} use the $3\times 2$pt data of DES-Y1 to constrain the fraction of decaying cold dark matter to $f_\mathrm{dcdm} < 0.037$, while the current strongest constraints, which come from the effective field theory of large scale structure applied to BOSS data, bound the fraction by $f_\mathrm{dcdm} < 0.0216$ if one assumes a lifetime smaller than the age of the Universe~\cite{Simon:2022ftd}. Lastly, references~\cite{DES:2020mpv, Schoneberg:2021qvd, Holm:2022eqq} find a difference in likelihood ratio of $\Delta \chi^2 \approx 0$ at the maximum-posterior point compared to $\Lambda$CDM, i.e. no preference for DCDM. In contrast, we find both two-sided bounds on $\Gamma$ and $f_\mathrm{dcdm}$ at $68 \%$ CL as well as an improved quality of fit to the data for DCDM using profile likelihoods instead of Bayesian inference.

\section{Profile likelihood}\label{sec:level3}
A profile likelihood (PL) is obtained by fixing a parameter of interest $\theta_0$ to different values and maximizing the likelihood $\mathcal{L}(\vec{\theta})$ with respect to all other parameters of the model, $\theta_1, \dots \theta_N$:
\begin{equation} \label{def:profile}
	\mathrm{PL}(\theta_0) =\min_{\theta_1,\dots,\theta_N} \Delta \chi^2 (\vec{\theta})
	= - 2 \ln \left(\max_{\theta_1,\dots \theta_N} \frac{\mathcal{L}(\vec{\theta})}{\mathcal{L}_\mathrm{max}}\right),
\end{equation}
where $\mathcal{L}_\mathrm{max}$ is the likelihood at the maximum likelihood estimate and $\vec{\theta}= (\theta_0,\dots, \theta_N)$.  Since the PL is only based on likelihood ratios, it is invariant under reparametrization of the parameter space $\vec{\theta}\rightarrow \vec{\theta}'$ as this does not change the maximum likelihood estimate. The confidence interval can then be obtained via the \textit{Neyman construction}~\cite{Neyman:1937uhy}: If $\theta_0$ follows a Gaussian distribution, i.e. the PL is parabolic, the confidence interval at confidence level $1-\alpha$ is given by the interval $\Delta\chi^2(\vec{\theta}) \le \chi^2_\alpha$, where $\chi^2_\alpha = 1$ (3.84) for $\alpha = 68\%$ ($95\%$); for example the $1\sigma$ confidence interval is given by the interval between the intersections of the PL with $\Delta\chi^2 =1$. Due to the reparameterization invariance of the PL, this procedure can also be applied to non-parabolic profiles inasmuch as there exists a reparameterization in which it is parabolic~\cite{planck2018}.

We construct the profile likelihoods by \emph{simulated annealing}~\cite{Kirkpatrick_1983, Hannestad:2000wx}, introducing a temperature parameter $T$ and modifying the likelihood according to $\mathcal{L} \rightarrow \mathcal{L}^{1/T}$.
Increasing the temperature flattens the likelihood landscape, while decreasing it enhances any peak structures. Simulated annealing works by running an MCMC chain while successively lowering the temperature, and generally performs well against noisy likelihood functions with many local maxima. We have implemented simulated annealing by running MCMC chains with the \textsc{MontePython} code~\cite{Audren:2012wb,Brinckmann:2018cvx} utilizing the Einstein--Boltzmann solver \textsc{class}~\cite{CLASS2} and supplying it with an approximate covariance matrix constructed from the MCMC results of~\cite{Nygaard:2020sow} to accommodate the difficult, high-dimensional likelihood landscape. We run each optimization several times to increase the confidence that the global maxima are being found, and assess the accuracy of our method to be on the level of $0.2$ in $\chi^2$. Our implementation is publicly available at \url{https://github.com/AarhusCosmology/montepython_public} on the branch whose name is the arXiv ID of this paper.

\section{Results}\label{sec:level4}

In this section, we present the results of our profile likelihood computations. Unless otherwise specified, we employ data from \textit{Planck} 2018 including high-$\ell$ TTTEEE, low-$\ell$ TT, EE and lensing~\cite{Planck_col_2020} as well as BAO data, including BOSS DR12~\cite{boss2016}, low redshift data from 6dF~\cite{beutler2011} and the BOSS main galaxy sample~\cite{ross2014}. Other than the DCDM parameters $\omega_\mathrm{dcdm}^\mathrm{ini}, \log_{10} \Gamma \ \mathrm{Gyr}^{-1}$, the free cosmological parameters we maximize over are $\{\omega_b, \omega_\mathrm{cdm}, H_0, \ln 10^{10} A_s, n_s, \tau_\mathrm{reio}\}$.
\begin{figure}[tb]
	\includegraphics[width=\columnwidth]{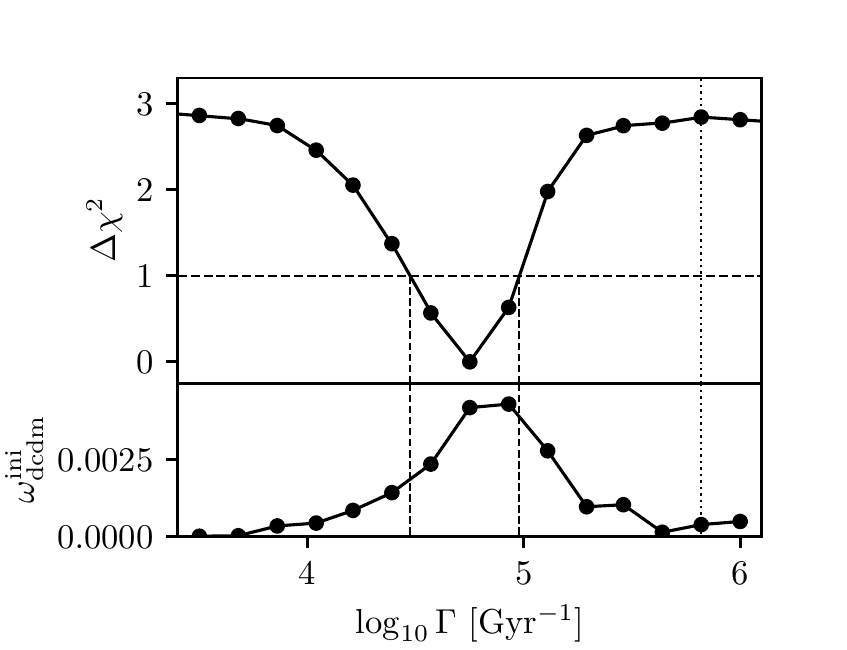}
	\caption{\label{fig:gamma_profile} Top panel: One-dimensional profile likelihood of $\log_{10} \Gamma/$Gyr, where $\Gamma$ denotes the decay constant of decaying dark matter, under data from Planck and BAO. Bottom panel: The abundance of decaying cold dark matter $\omega^{\mathrm{ini}}_\mathrm{dcdm}$ associated with every point in the profile, as obtained through the optimization. The flat regions at small and large $\Gamma$ correspond to the $\Lambda$CDM limit and the central well thus represents a $\Delta \chi^2 \approx 2.8$ improvement over $\Lambda$CDM. Dashed lines indicate the $\Delta \chi^2 = 1$ intersections, giving the $\approx 68 \%$ CIs.}
\end{figure}

Firstly, figure~\ref{fig:gamma_profile} shows a one-dimensional profile in the decay constant $\log_{10} \Gamma/$Gyr. The top panel shows the profile, while the bottom panel illustrates the decaying dark matter abundances $\omega^\mathrm{ini}_\mathrm{dcdm}$ associated with each point of the profile. Evidently, the model reduces to $\Lambda$CDM for $\log_{10} \Gamma/$Gyr $\lesssim 4$ and $\log_{10} \Gamma/$Gyr $\gtrsim 5.25$ due to the vanishing of $\omega^\mathrm{ini}_\mathrm{dcdm}$. Moreover, we have checked that the profile indeed continues to be flat for more extreme values of $\log_{10} \Gamma/$Gyr than shown here. Interestingly, in the intermediary regime, a well forms in $\Delta \chi^2$ associated with a non-zero decaying dark matter abundance. The maximum depth of the well is $\Delta \chi^2 \approx -2.8$, which corresponds to a $1.6\sigma$ preference for decaying dark matter over $\Lambda$CDM with two extra parameters. This is generally in contrast to the previous works~\cite{McCarthy:2022gok, Holm:2022eqq, Schoneberg:2021qvd} who all found $\Delta \chi^2 \approx 0.0$ when not including the SH0ES measurement. The approximate $68 \%$ CL bound obtained from the Neyman construction is $\log_{10} \Gamma \ \mathrm{Gyr}^{-1} = 4.763^{+0.214}_{-0.290}$, while we find that $\log_{10} \Gamma \ \mathrm{Gyr}^{-1}$ is unconstrained at $95 \%$ CL. This result contrasts the bounds obtained in previous Bayesian studies (e.g.~\cite{Poulin:2016nat, Nygaard:2020sow, Alvi:2022aam}), all of which either bound $\Gamma$ from above or below at $68 \%$ CL. The best-fit value $\log_{10} \Gamma/\mathrm{Gyr}=4.763$ corresponds to a decay just prior to recombination. We find this particularly interesting, since decaying cold dark matter now joins a class of many other $\Lambda$CDM extensions, such as early dark energy~\cite{Poulin:2018cxd, Herold:2021ksg, Herold:2022iib}, decaying warm dark matter/majorons~\cite{Holm:2022eqq, Escudero:2021rfi}, stepped dark radiation~\cite{Aloni:2021eaq, Schoneberg:2022grr} and variations of fundamental constants~\cite{Hart:2019dxi, Schoneberg:2021qvd}, all of which impact the physics exactly around recombination at their best-fit parameter values. The bestfit values of all parameters can be seen in table~\ref{bf_table}.
\begin{table}[tb]
	\begin{ruledtabular}
		\begin{tabular}{lcd}
			& $\Lambda$CDM & \text{DCDM} \\
			\colrule
			$100 \omega_b$                                   & $2.2362$  & $2.2358$ \\
			$\omega_\mathrm{cdm}$                            & $0.1191$  & $0.1209$ \\
			$H_0 \text{ [km s}^{-1}\text{ Mpc}^{-1}\text{]}$ & $67.71$   & $68.14$ \\
			$\ln (10^{10} A_s)$                              & $3.0491$  & $3.0585$ \\
			$n_s$                                            & $0.9684$  & $0.9767$ \\
			$\tau_\mathrm{reio}$                             & $0.05713$ & $0.05677$ \\
			$\omega_\mathrm{ini,dcdm}$                       & ---       & $0.00429$ \\
			$\log_{10} \Gamma_\mathrm{dcdm}/\mathrm{Gyr}$    & ---       & $4.763$ \\
			\colrule
			$\Omega_\Lambda$                                 & $0.6899$  & $0.6901$ \\
			$z_\mathrm{reio}$                                & $7.9676$  & $7.9677$ \\
			$\sigma_8$                                       & $0.8105$  & $0.8239$ \\
			$f_\mathrm{dcdm}$                                & ---       & $0.03428$ \\
		\end{tabular}
	\end{ruledtabular}
	\caption{\label{bf_table} Values of cosmological parameters at the bestfit of the decaying cold dark matter cosmology.}
\end{table}

Table~\ref{table} shows the likelihood budget at best-fit for the $\Lambda$CDM and decaying cold dark matter models. Evidently, the largest improvement is in the high-$\ell$ data which is very well-constrained by Planck, along with a mild improvement in low-$\ell$ TT data. Similar trends are seen in the $\chi^2$ budgets of the models that introduce new physics around recombination mentioned above~\cite{Schoneberg:2021qvd}.
\begin{table}[tb]
	\begin{ruledtabular}
		\begin{tabular}{lcdr}
			&
			$\chi^2 \textrm{(}\Lambda\textrm{CDM)}$ &
			\chi^2 \textrm{(DCDM)} &
			$\Delta \chi^2$ \\
			\colrule
			Planck high-$\ell$  & $2377.78$   & $2375.98$  & $-1.81$      \\
			Planck low-$\ell$ TT & $22.85$     & $21.92$    & $-0.93$     \\
			Planck low-$\ell$ EE & $396.62$    & $396.46$   & $-0.16$    \\
			Planck lensing       & $8.89$      & $9.04$     & $0.15$     \\
			BAO                  & $5.49$      & $5.43$     & $-0.06$    \\
			\colrule
			total                & $2811.64$   & $2805.37$  & $-2.81$     \\
		\end{tabular}
	\end{ruledtabular}
	\caption{\label{table} Breakdown of the individual contributions of different likelihoods to the $\chi^2$ budget of the best-fit cosmologies of the DCDM and $\Lambda$CDM models, respectively. Planck high-$\ell$ refers to the combined TTTEEE spectrum.}
\end{table}

\begin{figure}[tb]
	\includegraphics[width=\columnwidth]{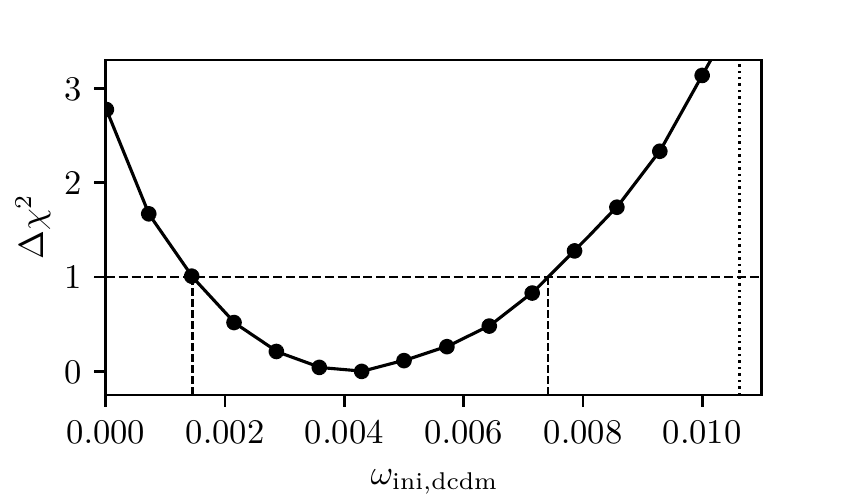}
	\caption{\label{fig:omega} Profile likelihood of the DCDM abundance $\omega^\mathrm{ini}_\mathrm{dcdm}$. Notably, the bestfit contains a non-zero DCDM component with $\chi^2_\mathrm{min} - \chi^2 (\omega^\mathrm{ini}_\mathrm{dcdm}=0) = -2.8$.}
\end{figure}

To assess the evidence for a non-zero DCDM component, we have computed a profile in the DCDM abundance $\omega_\mathrm{dcdm}^\mathrm{ini}$, shown in figure~\ref{fig:omega}. Interestingly, the maximum likelihood estimate is not in the $\Lambda$CDM limit, but at the intermediate value $\omega_\mathrm{dcdm}^\mathrm{ini} = 0.00429^{+0.00313}_{-0.00284}$, obtained from the approximate $68 \%$ CL Neyman construction,  corresponding to a bestfit $f_\mathrm{dcdm} =  0.034$. To our knowledge, no previous Bayesian analysis has found a lower bound on the DCDM abundance, and the difference with our result indicates a prior volume effect in the $\Lambda$CDM limit where $\omega_\mathrm{dcdm}^\mathrm{ini} \rightarrow 0$. The $\Lambda$CDM limit still lies within the $95 \%$ CI, however, and with the upper bound $\omega_\mathrm{dcdm}^\mathrm{ini} < 0.0106$, the decay constant $\Gamma$ is also entirely unconstrained at $95 \%$ CL.

\begin{figure}[tb]
	\includegraphics[width=\columnwidth]{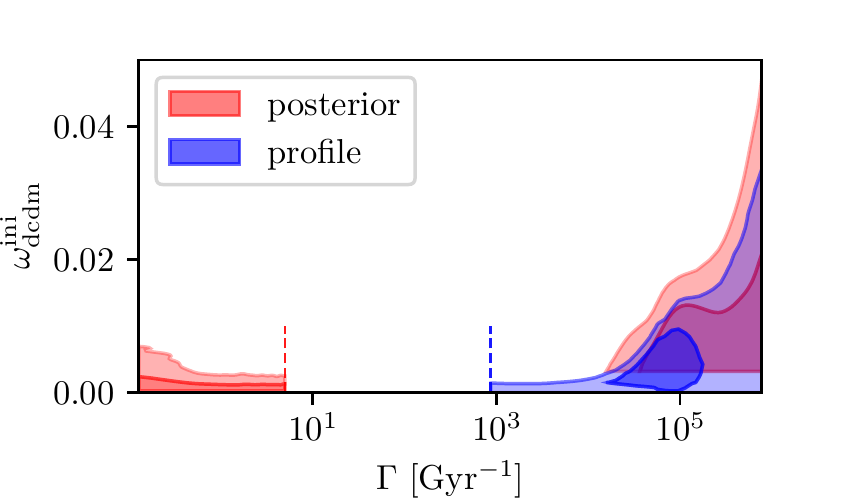}
	\caption{\label{fig:2d_profile} $1\sigma$ and $2\sigma$ contours of the decaying dark matter model parameters as estimated by the Bayesian posterior distribution (red) and a two-dimensional profile likelihood (blue). Dashed vertical lines represent bounds on the priors and on the sampling space in the former and latter case, respectively. The intermediate area has not been probed by MCMC methods in the literature, but we expect the two posterior patches to connect by some funnel of small $\omega_\mathrm{dcdm}^\mathrm{ini}$ values.}
\end{figure}

The reason for the difference between the MCMC and frequentist results is most clear from figure~\ref{fig:2d_profile}, where the $1\sigma$ and $2\sigma$ contours from a two-dimensional profile likelihood and Bayesian posterior distributions are compared. The posteriors of the short- and long-lived regimes are taken from~\cite{Nygaard:2020sow}, the former having a prior upper bound indicated by the dashed, vertical red line. The dashed, vertical blue line represents the lower bound of the grid on which we evaluated the profile likelihood, but we expect the flat behaviour to extrapolate. Evidently, the $\Lambda$CDM limit is always within the $1\sigma$ contour of the posterior, whereas it is outside the $1\sigma$ bound of the profile likelihood. The relaxation of the 2$\sigma$ profile and posterior at large $\Gamma$ is due to the model reducing to a model with extra relativistic degrees of freedom, $\Delta N_\mathrm{eff}$, and the distinct shape is a direct effect of parameterizing the abundance as $\omega^\mathrm{ini}_\mathrm{dcdm}$. In turn, marginalizing over $\omega_\mathrm{dcdm}^\mathrm{ini}$ creates the volume effect in the $\Gamma$ posterior that favours the largest possible values and makes the Bayesian inference highly prior dependent. Interestingly, a slight substructure can be seen in the posterior around the bestfit of the profile. We interpret the bump as a representation of the actual bestfit region, which is dragged and shrouded by the volume effects around the upper bound on $\Gamma = 10^6 \ \mathrm{Gyr}^{-1}$. Indeed, we have explicitly tested this by running additional MCMC estimations of the posterior, where it is seen that the bump is increasingly smoothed away as the chain converges, and we furthermore find a strong variation of the shape with the upper bound on $\Gamma$.

\begin{figure}[tb]
	\includegraphics[width=\columnwidth]{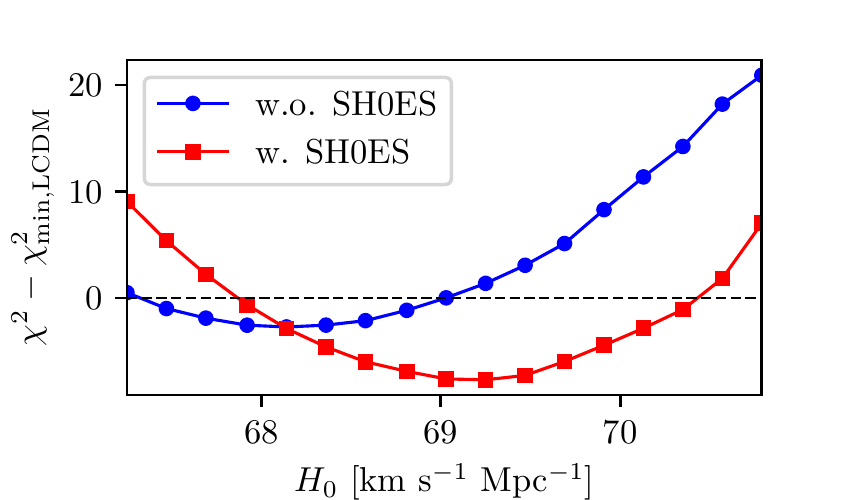}
	\caption{\label{fig:H0} Profile likelihood of $H_0$ with and without the SH0ES measurement of $H_0$, respectively.  The bestfit $\Lambda$CDM $\chi^2$ is subtacted in each case. Without SH0ES, the $H_0$ values are smaller than their MCMC equivalents and do not solve the $H_0$ tension.}
\end{figure}

Lastly, to assess the DCDM model's relation to the $H_0$ tension, we have computed profile likelihoods in $H_0$, shown in figure~\ref{fig:H0}, where the red profile includes a Gaussian likelihood on the $H_0$ measurement of~\cite{Riess_H0}. The second axis represents the difference in $\chi^2$ between the profiles and the best-fit $\Lambda$CDM models with and without the SH0ES measurement, respectively. In both cases, the DCDM model has a lower $\chi^2$ than $\Lambda$CDM for considerable intervals in $H_0$. Comparing the two, we see that the DCDM profile with the SH0ES measurement prefers a larger value for $H_0$ and has a larger difference in $\chi^2$ to the $\Lambda$CDM model, $\Delta \chi^2 = -7.8$, which, although a significant difference, is somewhat weaker than other well-performing $\Lambda$CDM extensions~\cite{Schoneberg:2021qvd}. With the Neyman construction, we find the bounds $H_0=69.25^{+0.32}_{-0.49} \ \mathrm{km} \ \mathrm{s}^{-1} \ \mathrm{Mpc}^{-1}$ and $H_0=68.14^{+0.54}_{-0.49} \ \mathrm{km} \ \mathrm{s}^{-1} \ \mathrm{Mpc}^{-1}$ at $68 \%$ CL with and without the SH0ES measurement, respectively. The latter corresponds to a mild alleviation of the $H_0$ tension from $4.1\sigma$ to $3.6\sigma$ using the Gaussian tension metric~\cite{Schoneberg:2021qvd}, valid here since the profiles are approximately quadratic. This is a stronger alleviation than that found with MCMC in references~\cite{Schoneberg:2021qvd, Simon:2022ftd} since the bestfit found here is outside of their prior choices, but actually a weaker alleviation than that found with MCMC in~\cite{Holm:2022eqq, Nygaard:2020sow}. Our results therefore corroborate the standing opinion in the literature that cold dark matter decaying to dark radiation cannot adequately solve the $H_0$ tension.

We finish this discussion by emphasising that results obtained from profile likelihoods are no more correct than ones obtained from an MCMC analysis. The two methods answer different questions: Bayesian methods localize bulk volumes in parameter space favoured by data, whereas profile likelihoods are sensitive to best-fit points, no matter how much probability mass is associated with them. They also have different weaknesses: Bayesian inference can be mislead by volume effects, and one can argue that profile likelihoods are susceptible to fine-tuning, i.e. very narrow best-fits that are poorly motivated physically. Hence, our view is that the two methods are complementary, and a fully nuanced analysis can only be obtained by including both.

\section{Conclusion}\label{sec:level5}
In this work, we illustrated that Bayesian studies of the DCDM model are influenced by parameter volume effects in the DCDM abundance $\omega_\mathrm{dcdm}^\mathrm{ini}$ and decay constant $\Gamma$. Whereas several earlier works either bound $\Gamma$ from above or below, depending on the choice of prior, we find $\log_{10} \Gamma \ \mathrm{Gyr}^{-1} = 4.763^{+0.214}_{-0.290}$ at $68 \%$ CL using the prior-independent profile likelihood, here with \textit{Planck} 2018 high-$\ell$ TTTEEE, low-$\ell$ TT, EE, lensing and BAO data. Interestingly, these values correspond to decays occurring around recombination. Moreover, we find the first (to our knowledge) $68 \%$ CL lower bound on the DCDM abundance, $\omega_\mathrm{dcdm}^\mathrm{ini} = 0.00429^{+0.00313}_{-0.00284}$. At $95 \%$ CL, however, we find only an upper bound on $\omega_\mathrm{dcdm}^\mathrm{ini}$ and thus an unconstrained lifetime. 

The relative preference over $\Lambda$CDM is around $\Delta \chi^2 \approx -2.8$ and $\Delta \chi^2 \approx -7.8$ with and without the SH0ES $H_0$ measurement, respectively. The former is similar to other popular $\Lambda$CDM extensions such as early dark energy or majorons, but the latter is a significantly weaker improvement than many of these models. This can be attributed to the inability of DCDM to relieve the $H_0$ tension, as seen from our profile likelihoods in $H_0$ with the bound $H_0 = 68.14^{+0.54}_{-0.49}$ km s$^{-1}$ Mpc$^{-1}$. In conclusion, our work shows that the bestfitting DCDM parameters correspond not to the stable or short-lived scenarios, but to a decay of around $3 \%$ of cold dark matter around recombination. Although unable to solve the $H_0$ tension, the DCDM model appears more viable than previously assumed.

\section*{Acknowledgements}
We thank Elisa Ferreira, Eiichiro Komatsu and Julien Lesgourgues for valuable discussions and comments. We acknowledge computing resources from the Centre for Scientific Computing Aarhus (CSCAA). E.B.H, A.N. and T.T. were supported by a research grant (29337) from VILLUM FONDEN.

\bibliographystyle{utcaps}
\bibliography{paper}

\end{document}